\newcommand{\al}{\mbox{$^{26}$\hspace{-0.2em}Al}}
\newcommand{\fe}{\mbox{$^{60}$Fe}}
\newcommand{\Msol}{\mbox{$M_{\sun}$}}

\newcommand{\sun}{\hbox{$\odot$}}
\def\MeV{\mbox{Me\hspace{-0.1em}V}}

\documentstyle[epsfig]{aipproc}
\begin{document}

\title{Gamma-ray line emission from OB associations and young open clusters}
\author{J\"urgen Kn\"odlseder$^{\ast,\dagger}$, 
        Miguel Cervi\~no$^{\ddagger}$, 
        Daniel Schaerer$^{\ddagger}$ and 
        Peter von Ballmoos$^{\dagger}$}
\address{$^{\ast}$INTEGRAL Science Data Centre, Chemin d'Ecogia 
         16, 1290 Versoix, Switzerland\\
         $^{\dagger}$Centre d'Etude Spatiale des Rayonnements, B.P.~4346, 31028 Toulouse 
         Cedex 4, France\\
         $^{\ddagger}$Observatoire Midi-Pyr\'en\'ees, 14, avenue Edouard Belin,
         31400 Toulouse, France}

\maketitle

\begin{abstract}
OB associations and young open clusters constitute the most prolific 
nucleosynthesis sites in our Galaxy.
The combined activity of stellar winds and core-collapse supernovae 
ejects significant amounts of freshly synthesised nuclei into the 
interstellar medium.
Radioactive isotopes, such as \al\ or \fe, that have been 
co-produced in such events may eventually be observed by gamma-ray 
instruments through their characteristic decay-line signatures.
However, due to the sensitivity and angular resolution of current (and 
even future) $\gamma$-ray telescopes, only integrated $\gamma$-ray line 
signatures are expected for massive star associations.

In order to study such signatures and to derive constraints on 
the involved nucleosynthesis processes, we developed a multi-wavelength 
evolutionary synthesis model for massive star associations.
This model combines latest stellar evolutionary tracks and 
nucleosynthesis calculations with atmosphere models in order to 
predict the multi-wavelength luminosity of a given association as 
function of its age.

We apply this model to associations and clusters in the well-studied Cygnus 
region for which we re-determined the stellar census based on 
photometric and spectroscopic data.
In particular we study the relation between 1.809 \MeV\ $\gamma$-ray line 
emission and ionising flux, since the latter has turned out to provide an 
excellent tracer of the global galactic 1.809 \MeV\ emission.
We compare our model to COMPTEL 1.8 \MeV\ $\gamma$-ray line 
observations from which we derive limits on the relative contributions from 
massive stars and core-collapse supernovae to the actual \al\ content 
in this region.
Based on our model we make predictions about the expected \al\ 
and \fe\ line signatures in the Cygnus region. 
These predictions make the Cygnus region a prime target for the future 
INTEGRAL mission.
\end{abstract}

\section*{Evolutionary Synthesis Model}

Our evolutionary synthesis model is based on the multi-wavelength code 
described in \cite{mashesse91,cervino94}, enhanced by the inclusion of 
nucleosynthesis yields.
In summary, the evolution of each individual star in a stellar population 
is followed using Geneva evolutionary tracks with enhanced 
mass-loss rates \cite{meynet94}.
In our present implementation, and similar to \cite{mashesse91,cervino94}, 
stellar Lyman continuum luminosities are taken from \cite{mihalas72,kurucz79}.
Note, however, that modern atmosphere models including the effects of line 
blanketing and stellar winds predict enhanced ionising fluxes with 
respect to these models, hence our predicted ionising luminosities should be 
considered as preliminary and possibly are somewhat too low\footnote{We currently 
are implementing the CoStar atmosphere models of \cite{schaerer97} in our code 
that consistently treat the stellar structure and atmosphere and include line 
blanketing and stellar winds.}.
At the end of stellar evolution, stars initially more massive than 
$M_{\rm WR}=25$ \Msol\ are exploded as Type Ib supernovae, while stars 
of initial mass within $8\Msol$ and $M_{\rm WR}$ are assumed to 
explode as Type II SNe.
Nucleosynthesis yields have been taken from \cite{meynet97} for the 
pre-supernova evolution and from \cite{woosley95,wlw95} for Type II 
and Type Ib supernova explosions, respectively.
Note that Type II SN yields have only been published for stars without 
mass loss and Type Ib yields have only been calculated for pure Helium 
stars.
In order to obtain consistent nucleosynthesis yields for Type II 
supernovae we followed the suggestion of \cite{maeder92} and linked 
the explosive nucleosynthesis models of \cite{woosley95} to the Geneva 
tracks via the core mass at the beginning of Carbon burning.
For Type Ib SN we used the core mass at the beginning of He core 
burning to link evolutionary tracks to nucleosynthesis calculations.

\begin{figure}[t!]
\epsfig{file=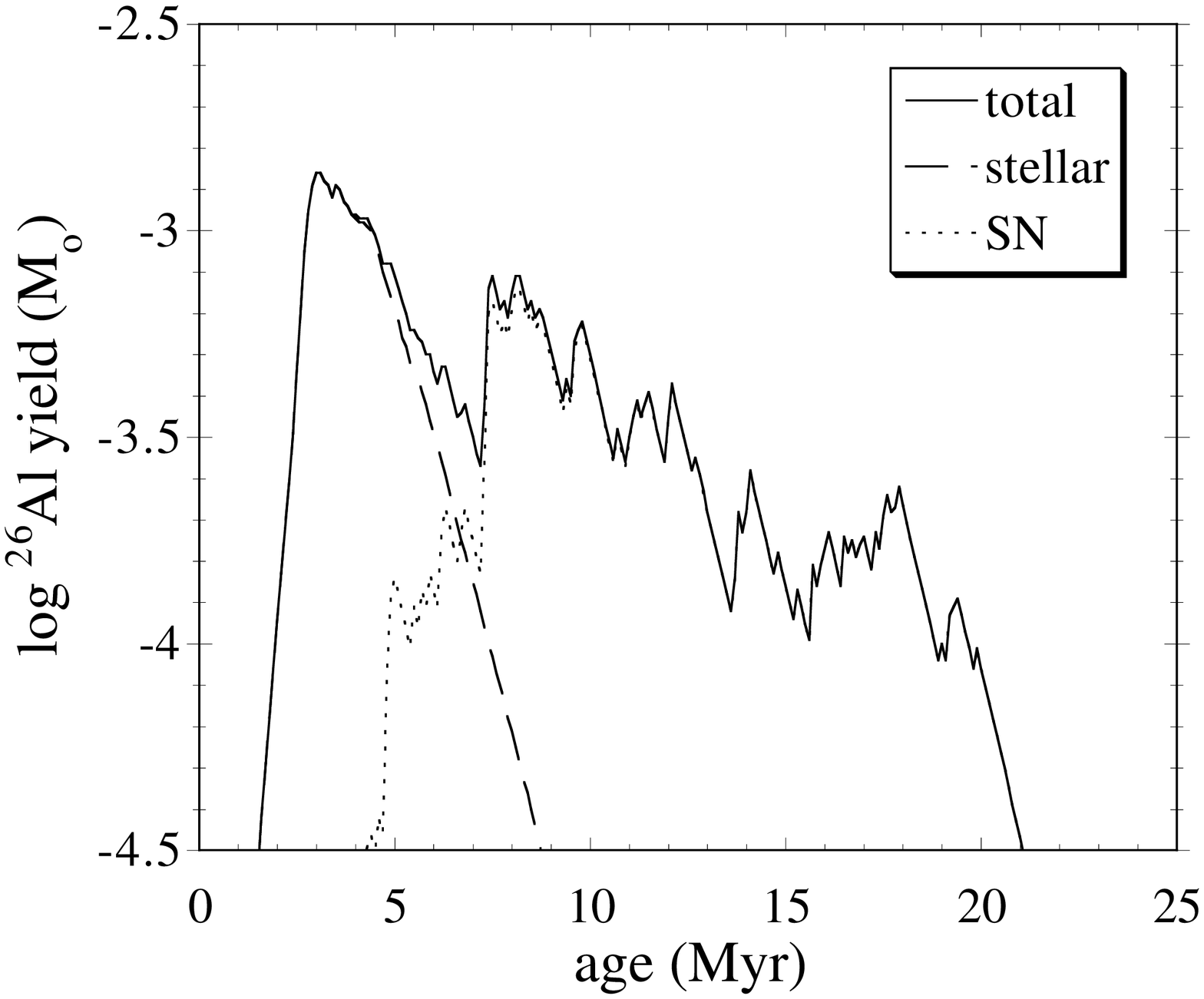,width=7.3cm}
\hfill
\epsfig{file=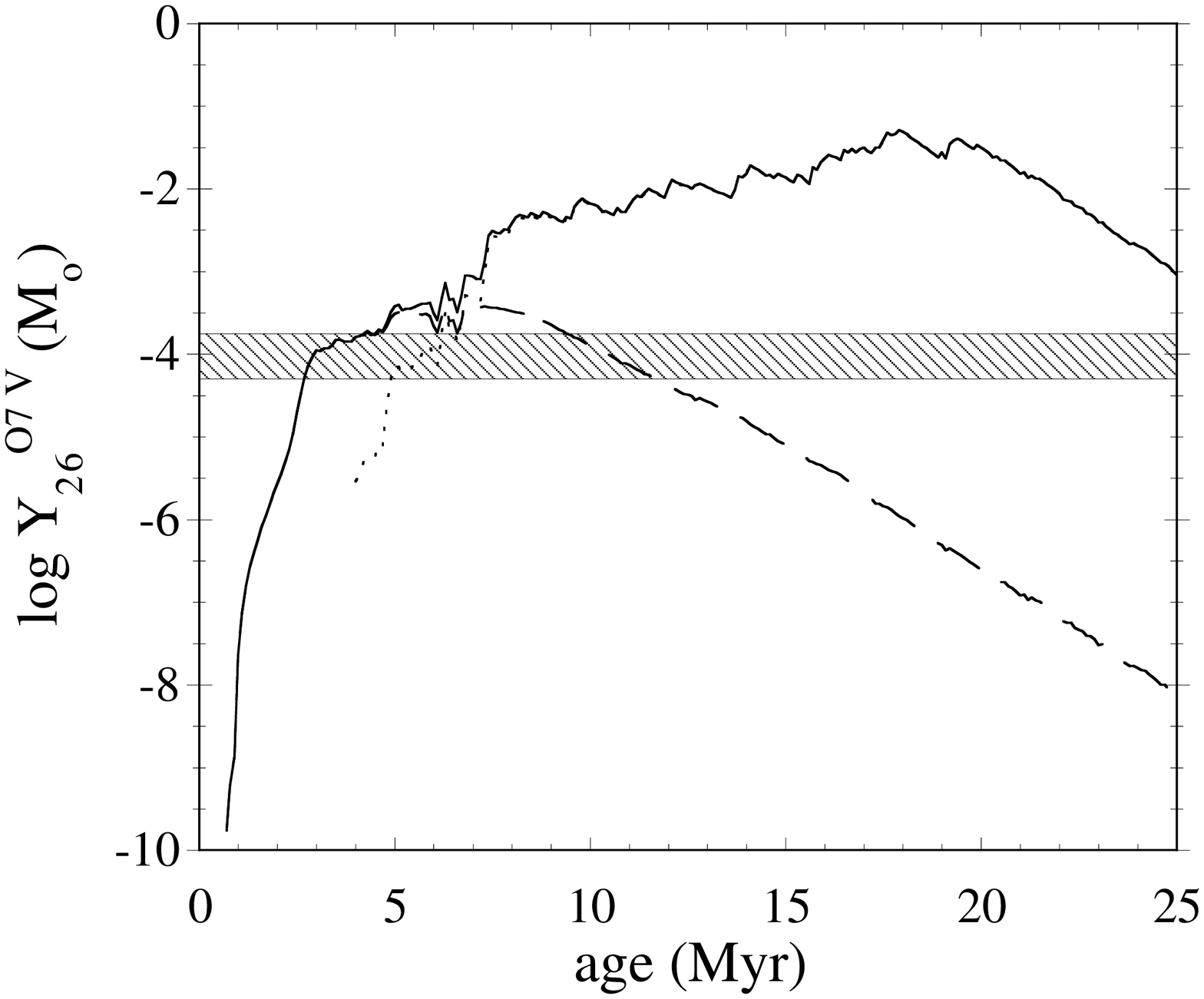,width=7.3cm}
\vspace{10pt}
\caption{\al\ yield (left) and equivalent O7V star \al\ yield (right) as function 
of association age. The hatched area indicates the 
COMPTEL measurement of $Y_{26}^{\rm O7 V} = (1.0 \pm 0.3)\,10^{-4}$ \Msol.}
\label{fig:al26}
\end{figure}

Evolutionary synthesis models were calculated using a stochastic 
initial mass function where random masses were assigned to 
individual stars following a Salpeter initial mass spectrum 
($d \log \xi / d\log M = -1.35$) until the number of stars in a given 
mass-interval reproduces the observed population.
Typical results for a rich OB association (51 stars within $15-40$ 
\Msol) are shown in Fig.~\ref{fig:al26}.
\al\ production turns on at about $1-2$ Myr when the isotope starts to 
be expelled by stellar winds into the interstellar medium.
Stellar \al\ production reaches its maximum around 3 Myr when the 
most massive stars enter the Wolf-Rayet phase.
Explosive nucleosynthesis sets on around 4.5 Myr for Type Ib and around 7 Myr 
for Type II supernovae, leading to a second peak in the \al\ yield 
around 8 Myr.
After this peak, a slightly declining \al\ yield is maintained by Type 
II explosions until the last Type II SN exploded around 20 Myr.
Afterwards, the exponential decay quickly removes the remaining \al\ nuclei 
from the ISM.

We also calculated the time-dependent ionising luminosity ($\lambda < 912$ {\AA}) 
of the population from which we derived the equivalent O7V star \al\ yield 
$Y_{26}^{\rm O7 V}$.
This quantity measures the amount of \al\ ejected per ionising photon normalised 
on the ionisation power of an O7V star.
The analysis of COMPTEL 1.809 \MeV\ data suggests a galaxywide 
constant value of $Y_{26}^{\rm O7V} = (1.0 \pm 0.3)\,10^{-4}$ \Msol\
\cite{knoedl99}.
Interestingly, the COMPTEL value is only reproduced for a very young 
population during a quite short age period ($2.5-5$ Myrs).
For younger populations too few \al\ is produced with respect to the 
ionising luminosity, leading to much lower equivalent yields.
For older populations the ionising luminosity drops rapidly, resulting
in much higher $Y_{26}^{\rm O7V}$ values.
Thus, the equivalent O7V star \al\ yield is a quite sensitive 
measure of the population age.
In particular, the measurement of $Y_{26}^{\rm O7V}$ for individual OB 
associations or young open clusters provides a powerful tool to 
identify the dominant \al\ progenitors.

\section*{Application to the Cygnus region}

We applied our evolutionary synthesis model to the Cygnus region from 
which prominent 1.809 \MeV\ line emission has been detected by COMPTEL
\cite{delRio96}.
In \cite{delRio96} the authors modelled \al\ nucleosynthesis in 
Cygnus by estimating the contribution from individual Wolf-Rayet stars 
and supernova remnants that are observed in this region.
This approach, however, suffers from considerable uncertainties due 
to the poorly known distances to these objects.
In this work we performed a complete census of OB associations and 
young open clusters in the Cygnus region.
Individual association or cluster distances have been estimated by the method 
of spectroscopic parallaxes, ages have been determined by isochrone fitting.
Distance and age uncertainties have also been estimated and were
incorporated in the analysis by means of a Bayesian method.
The richness was estimated for each association or cluster by 
building H-R diagrams for member stars and by counting the number of 
stars within mass-intervals that are probably not affected by 
incompleteness or evolutionary effects.
In total we included 6 OB associations and 19 young open clusters in 
our analysis which house 94 O type and 13 Wolf-Rayet stars.

For each OB association or cluster, 100 independent evolutionary 
synthesis models were calculated that differ by the actual stellar 
population that has been realised by the random sampling procedure.
In this way we include the uncertainties about the unknown number of 
massive stars in the associations that have already disappeared in 
supernova explosions.
From these samples the actual age and distance uncertainties are 
eliminated by marginalisation, leading to a probability distribution 
for all quantities of interest.
Note that in this approach an age uncertainty is equivalent to an 
age spread in the cluster formation, hence the possibility of 
non-instantaneous star formation has been taken into account.
The results for all individual associations have been combined by 
marginalisation to predictions for the entire Cygnus region.

The predicted equivalent O7V star \al\ yield amounts to 
$(0.3-1.2)\,10^{-4}$ \Msol\ and is compatible with the COMPTEL 
observation, pointing to an extremely young population that is at the 
origin of \al.
Indeed, while $90\%$ of the \al\ is produced in our model by stellar 
nucleosynthesis (during the main sequence and subsequent Wolf-Rayet 
phase), only $10\%$ may be attributed to explosive 
nucleosynthesis, mainly in Type Ib SN events.
This is also reflected in the low \fe\ yields ($(0-7)\,10^{-8}$ ph 
cm$^{-2}$ s$^{-1}$) that are predicted by our model since \fe\ is assumed 
to be only produced in supernovae.

However, in absolute quantities, our model underestimates the free-free 
intensity in Cygnus by about a factor 3 while the total \al\ flux is 
low by a factor of 5.
This points towards a possible incompleteness of our massive star 
census which has been based on surveys of OB associations and young open 
clusters in Cygnus available in the literature.
Indeed, while we identify only 95 OB stars in Cyg OB2, \cite{reddish66}
estimated 400 OB members in this association, indicating only $25\%$ 
completeness of our census.
Taking $25\%$ as a typical completeness fraction for our OB association 
census and assuming that the young open cluster census is complete,
we obtain a free-free intensity of $0.25$ mK and an \al\ flux of 
$4.3\,10^{-5}$ ph cm$^{-2}$ s$^{-1}$ -- values that are in fairly 
good agreement with the observations ($0.26$ mK from DMR microwave 
data and $(7.9\pm2.4)\,10^{-5}$ ph cm$^{-2}$ s$^{-1}$ from COMPTEL 
1.8 \MeV\ observations; see Pl\"uschke et al., these proceedings).
However, we do not predict any noticeable amount of \fe\ for the 
Cygnus region -- a prediction which hopefully will be soon verified by 
the INTEGRAL observatory.
We would like to stress that INTEGRAL has the potential to partially 
resolve some OB associations and young open clusters in the nearby Cygnus 
region, and thus may provide important new insights in massive star 
nucleosynthesis in this area.



\end{document}